# Selected Multi-Criteria Decision-Making Methods and Their Applications to Product and System Design

**Zhiyuan Wang[1, 2], Seyed Reza Nabavi[3] and Gade Pandu Rangaiah[1, 4]**

1. Department of Chemical and Biomolecular Engineering, National University of Singapore, Singapore 117585, Singapore
2. Department of Computer Science, DigiPen Institute of Technology Singapore, Singapore 139660, Singapore
3. Department of Applied Chemistry, Faculty of Chemistry, University of Mazandaran, Babolsar, Iran
4. School of Chemical Engineering, Vellore Institute of Technology, Vellore 632014, India

## 15.1 Overview

Optimization has found numerous applications in engineering, particularly since 1960's. Many optimization applications in engineering have more than one objective (or performance criterion). Such applications require multi-objective (or multi-criteria) optimization (MOO or MCO). Spurred by this and development of techniques for handling multiple objectives, MOO has found many applications in engineering in the last two decades. Optimization of an application for more than one objective gives a set of optimal solutions (known as non-dominated or Pareto-optimal solutions), which are equally good in the sense that no objective can be further improved without resulting in deterioration of at least one other objective. MOO in engineering has mainly focused on development of a model for the application, formulation of the MOO problem and solution of the formulated problem to find Pareto-optimal solutions. However, for completion of MOO, one more step is required to choose one of these optimal solutions for implementation.

Many methods are available for ranking non-dominated solutions and selecting one of them. They are known as multi-criteria decision making or analysis (MCDM or MCDA) methods. This step of using MCDM methods has not received much attention in engineering applications of MOO. In this chapter, selected MCDM methods are described, and then applied to several applications in product and system design. First, generic procedure of MCDM is presented in Section 15.2. Normalization and weighting methods, required for many MCDM methods, are described in Section 15.3. Then, selected MCDM methods are explained in Section 15.4. A Microsoft Excel program, EMCDM445, for these methods and its use are outlined in Section 15.5. Application of selected MCDM methods to example datasets on product and system design and discussion of its results are covered in Section 15.6. Finally, this chapter ends with a summary in Section 15.7.

Learning outcomes of this chapter on MCDM methods are:
1. State the need for MCDM methods
2. Outline the generic procedure of MCDM methods
3. Explain normalization and weighting methods
4. Describe selected MCDM methods
5. Apply MCDM methods to the given dataset of Pareto-optimal solutions
6. Interpret the results from the application of MCDM methods

This chapter will be useful to those readers new to MCDM whereas the EMCDM445 program for selected MCDM methods will be handy to all readers in their study of MOO and MCDM.

## 15.2 Procedure of MCDM

The **five steps** of MCDM are presented in Figure 15.1. The **first step** is to start with a set of non-dominated solutions, which are found by formulating and solving the MOO problem under study. In the MOO problem, each objective/criterion can be for either maximization (also referred to as the benefit objective) or minimization (also referred to as the cost objective). Instead of solving the MOO problem, available alternatives (or solutions) and their performance criteria can be compiled and used in MCDM.

The **second step** is to construct an objective matrix, which is also called as decision matrix in some publications. This matrix consists of $m$ rows (with one row for each non-dominated solution or alternative) and $n$ columns (with one column for each objective or criterion). Hence, the objective/decision matrix does not contain values of decision variables or constraints in the MOO problem.

In the **third step**, the objective values are normalized using one of the normalization methods, such as vector and max-min normalization methods. One can choose to skip the normalization step (e.g., when values of all objectives are comparable). However, normalization is recommended to avoid domination of MCDM computations by those objectives with relatively large values compared to other objectives, which is the situation in many engineering applications.

In the **fourth step**, weight for each of the objectives is provided (e.g., by decision makers) or computed using one of the weighting methods such as the entropy method. Equal weights for all objectives, if appropriate, can be given. The **fifth step** is to rank the non-dominated solutions using one or several MCDM methods, which may rank the solutions differently. In the **sixth and final** step, review the 3 or 4 top-ranked solutions by one or more MCDM methods, for selecting one of them for implementation. In this selection, optimal values of decision variables in the MOO problem, if available, or other considerations may be used.

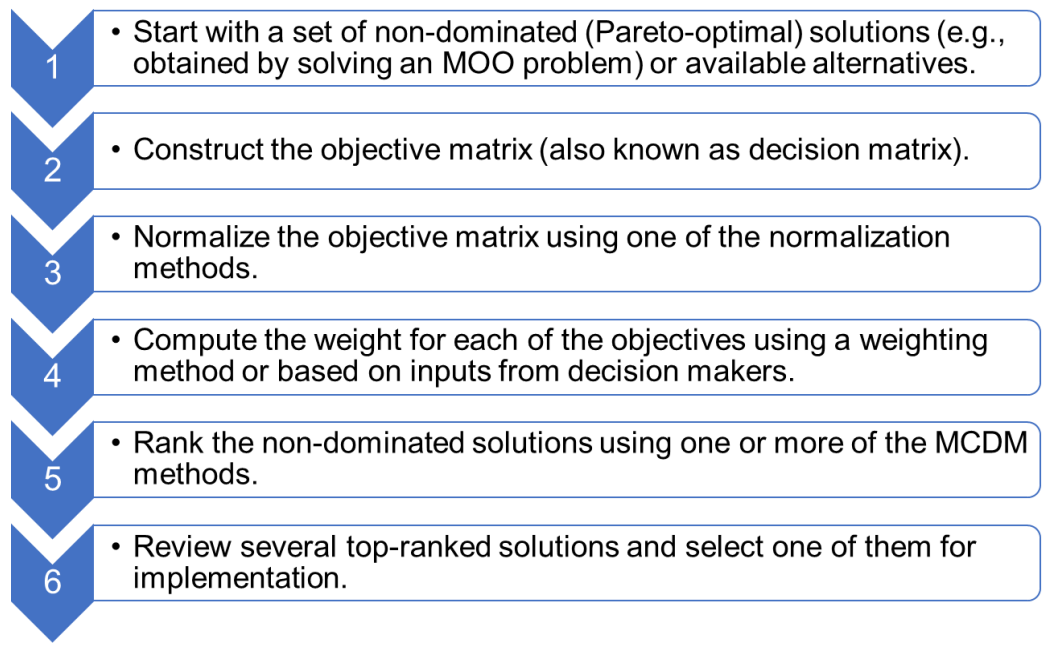

Figure 15.1: General steps of MCDM.

**15.3   Normalization and Weighting Methods**

For describing the popular normalization, weighting methods and MCDM methods (which are all implemented in the Excel-based MCDM program, described in Section 15.5), together with their mathematical equations in this and next sections, the following symbols are introduced, and they are used throughout this chapter. The objective matrix is a matrix of $m$ rows (with one row for each non-dominated solution or alternative) and $n$ columns (with one column for each objective or criterion). $f_{ij}$ represents the value of the $j$th objective at the $i$th non-dominated (Pareto-optimal) solution, $F_{ij}$ represents the normalized $f_{ij}$ after applying one of the normalization methods, and $w_j$ is the weight of the $j$th objective. Besides, note that the range of integers $i$ and $j$ are $i \in [1, \ m]$ and $j \in [1, \ n]$, respectively.

Normalization is an important step in bringing all values in the objective matrix to a common scale (e.g., between 0 and 1). This is effective to prevent objectives with relatively large values dominating the calculations in MCDM. Table 15.1 presents the four normalization methods, namely, sum, vector, max-min and max normalization, and their mathematical equations. For the range of normalized values, assuming all $f_{ij}$ values are greater than 0, two cases may be used to examine the normalized values, $F_{ij}$: (i) range of values of an objective is narrow, e.g., within 5%, and (ii) ratio of maximum to minimum of an objective is $\mu$. In case (i), normalized values by sum and vector normalizations will be approximately $1/m$ and $1/\sqrt{m}$, and normalized values by max normalization will be close to 1.0. In case (ii), on the other hand, normalized values by sum normalization will be $\left[\frac{1}{1+\mu(m-1)}, \frac{\mu}{\mu+(m-1)}\right]$, by vector normalizations will be $\left[\frac{1}{\sqrt{1+\mu^2(m-1)}}, \frac{\mu}{\sqrt{\mu^2+(m-1)}}\right]$, and by max normalization will be $\left[\frac{1}{\mu}, 1.0\right]$. In both cases (i) and (ii), normalized values by max-min normalization will be between 0.0 and 1.0.

After normalization by either max-min or max method, normalized values of a cost objective will be like those of a benefit objective (i.e., higher values are preferred). Note that division by zero may occur while using sum and max normalizations (Table 15.1): in sum normalization, when an objective has some positive and some negative values, and their sum is zero; and in max normalization, when the maximum value of a benefit objective is zero or any value of a cost objective is zero. Further, all values of a cost objective by max normalization become zero if the minimum of that cost objective is zero. Although these situations are not common, care should be exercised if the value of an objective is zero or values of an objective include both positive and negative values.

Table 15.1: Normalization methods and their mathematical equations

| Method | Equations, where $i \in [1, m]$ and $j \in [1, n]$ |
|---|---|
| Sum Normalization | $F_{ij} = \dfrac{f_{ij}}{\sum_{k=1}^{m} f_{kj}}$ for both benefit and cost objectives |
| Vector Normalization | $F_{ij} = \dfrac{f_{ij}}{\sqrt{\sum_{k=1}^{m} f_{kj}^2}}$ for both benefit and cost objectives |
| Max-min Normalization | $F_{ij} = \dfrac{f_{ij} - min_{k\in[1,m]} f_{kj}}{max_{k\in[1,m]} f_{kj} - min_{k\in[1,m]} f_{kj}}$ for benefit objectives <br><br> $F_{ij} = \dfrac{max_{k\in[1,m]} f_{kj} - f_{ij}}{max_{k\in[1,m]} f_{kj} - min_{k\in[1,m]} f_{kj}}$ for cost objectives |
| Max Normalization | $F_{ij} = \dfrac{f_{ij}}{max_{k\in[1,m]} f_{kj}}$ for benefit objectives <br><br> $F_{ij} = \dfrac{min_{k\in[1,m]} f_{kj}}{f_{ij}}$ for cost objectives |

There are many methods for calculating the weight of each of $m$ objectives. Some of them compute weights using the values in the objective matrix and do not require any inputs from the user or decision maker, and they are known as objective weighting methods; in this phrase, objective is the adjective, and it means that not influenced by personal feelings or opinions in considering and representing facts. Some other weighting methods require inputs from the user or decision maker (and not the objective matrix) for calculating weights, and they are known as subjective weighting methods. Inputs can vary from one decision maker to another, and consequently the computed weights are subjective. In the following, two objective weighting methods and two subjective weighting methods are described. See Wang et al. (2020) for some more objective and subjective weighting methods.

**Entropy weighting method** (Hwang and Yoon, 1981) relies on a probability-theoretic formulation of a measure of informational uncertainty to decide the weights for objectives. An objective is allocated a relatively higher weight if there is a more significant difference in the values of that objective among the non-dominated solutions. According to the review by Hafezalkotob et al. (2019), entropy method is the most frequently used weighting method in MCDM applications. It consists of the following three steps.

**Step <1>:** Normalize the objective matrix using the sum normalization (Table 15.1), as in Hwang and Yoon (1981).

**Step <2>:** Compute the entropy value of each objective in the normalized objective matrix:

$$E_j = -\frac{1}{\ln(m)}\sum_{i=1}^{m}(F_{ij} \ln F_{ij}) \quad j \in [1, n] \tag{15.1}$$

Note that the above equation uses normalized objective values.

**Step <3>:** Find the weight of each objective based on computed entropy values:

$$w_j = \frac{1-E_j}{\sum_{j=1}^{n}(1-E_j)} \quad j \in [1, n] \tag{15.2}$$

Although sum normalization is used in Step 1 as per Hwang and Yoon (1981), normalization method employed affects weights calculated by the entropy method; see Wang et al. (2020) for some examples.

Another weighting method, namely, **Criteria Importance Through Intercriteria Correlation (CRITIC)** is based on the correlation between any two objectives and the standard deviation of each individual objective. The original study by Diakoulaki et al. (1995) reveals that this method produces more evenly distributed weights that consider the information offered by all objectives and enable a better resolution of trade-offs among the objectives. It also has three steps as follows. Like the entropy method, weights given by CRITIC method depend on the normalization method used in Step 1 (Wang et al., 2020).

**Step <1>:** Normalize the objective matrix using the max-min normalization method (Table 15.1), as in Diakoulaki et al. (1995).

**Step <2>:** Calculate the Pearson product moment correlation between any 2 objectives:

$$\rho_{jk} = \frac{\sum_{i=1}^{m}(F_{ij}-\overline{F_j})(F_{ik}-\overline{F_k})}{\sqrt{\sum_{i=1}^{m}(F_{ij}-\overline{F_j})^2}\sqrt{\sum_{i=1}^{m}(F_{ik}-\overline{F_k})^2}} \quad j,k \in [1, n] \tag{15.3}$$

Here, $\overline{F_j} = \frac{1}{m}\sum_{i=1}^{m} F_{ij}$ and $\overline{F_k} = \frac{1}{m}\sum_{i=1}^{m} F_{ik}$ represent the arithmetic mean of $j$ th and $k$ th normalized objective, respectively.

**Step <3>.** Compute the standard deviation of each individual normalized objective and then determine the weight for each objective as follows:

$$\sigma_j = \sqrt{\frac{\sum_{i=1}^{m}(F_{ij}-\overline{F_j})^2}{m}} \quad j \in [1, n] \tag{15.4}$$

$$c_j = \sigma_j \sum_{k=1}^{n}(1-\rho_{jk}) \quad j \in [1, n] \tag{15.5}$$

$$w_j = \frac{c_j}{\sum_{k=1}^{n} c_k} \quad j \in [1, n] \tag{15.6}$$

**Analytical Hierarchy Process (AHP)**, developed by Saaty (1990), is a popular subjective weighting method in MCDM applications (Hafezalkotob et al., 2019). It involves a sequence of pair-wise comparisons among the objectives for generating the objective weights. It does not require any values from the objective matrix; instead, it solely leverages the decision makers' subjective preferences to determine the weights for objectives. AHP is conveniently described in three steps, using a simple example.

**Step <1>:** Consider an objective matrix with 3 objectives, namely, Obj_A, Obj_B, and Obj_C (i.e., *n* = 3). The first pair-wise comparison given by decision makers in top row of Figure 15.2 shows that the favor (preference) level of Obj_B over Obj_A is 3 on the scale of from -9 to +9 (where -9 indicates the Obj_B is least favored, +9 indicates Obj_B is most favored, and 0 indicates equally favoring of Obj_A and Obj_B). Likewise, for the second pair-wise comparison of Obj_C over Obj_A, decision makers consider that Obj_C is less favored than Obj_A and hence give a -5. For the third pair-wise comparison of Obj_C over Obj_B, decision makers consider that Obj_C is more favored than Obj_B and hence give a 7. Note that decision makers are not confined to choose from only the 9 integer values on the scale; they can choose any other values within the range of [-9, 9], such as +2.5, +4.7 and -3.4. For *n* objectives, there will be $\frac{n(n-1)}{2}$ pair-wise comparisons; hence, decision makers may have to provide many inputs.

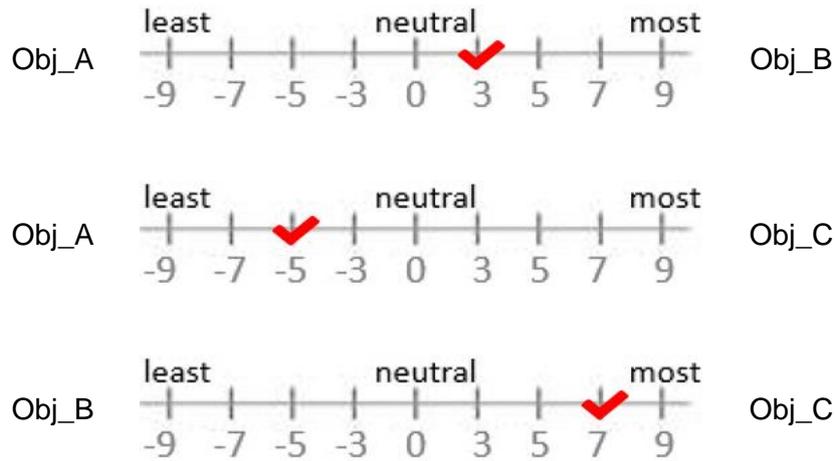

Figure 15.2: Pair-wise comparisons of 3 objectives in AHP method; each tick symbol shows the favor (preference) level given by decision makers for the comparison of the right-side objective over the left-side objective.

**Step <2>:** Construct the pair-wise comparison matrix illustrated in Table 15.2. All values on the leading diagonal of the matrix are set as 1. For all values above the leading diagonal, if the corresponding favor level given by the decision maker is positive (> 0), then take its reciprocal value in the matrix; if the corresponding favor level given by the decision maker is negative (< 0), then use its absolute value in the matrix. If the corresponding favor level given by the decision maker is zero (i.e., neutral), it will be taken as 1, which is same as the unity value on the diagonal. Values below the leading diagonal are reciprocal of the corresponding values above the leading diagonal (i.e., $a_{gj} = \frac{1}{a_{jg}}$ for $j \in [1,\ n-1]$ and $g \in [j+1,\ n]$; here, $a_{gj}$ is the value at the $g$th row and $j$th column of the matrix. For the simple example, the complete pair-wise comparison matrix based on inputs in Figure 15.2 is presented in Table 15.2.

Table 15.2: Pair-wise comparison matrix of 3 objectives, for AHP method

|       | Obj_A | Obj_B | Obj_C |
|-------|-------|-------|-------|
| Obj_A | 1     | 1/3   | 5     |
| Obj_B | 3     | 1     | 1/7   |
| Obj_C | 1/5   | 7     | 1     |

**Step <3>:** Apply sum normalization to the pair-wise comparison matrix constructed in the previous step. Here, sum is based on each column.

$$A_{gj} = \frac{a_{gj}}{\sum_{h=1}^{n} a_{hj}} \qquad g,j \in [1, \ n] \tag{15.7}$$

Then, weights for the objectives are computed as

$$w_g = \frac{\sum_{j=1}^{n} A_{gj}}{n} \qquad g \in [1, \ n] \tag{15.8}$$

Table 15.3 shows the normalized pair-wise comparison matrix (computed by equation 15.7) and weights for the objectives (computed by equation 15.8), for the example.

Table 15.3: Normalized pair-wise comparison matrix and weights for the example, by AHP method

|  | Pair-wise Comparison Matrix | | | Weight |
| --- | --- | --- | --- | --- |
|  | Obj_A | Obj_B | Obj_C |  |
| Obj_A | 0.238095 | 0.040000 | 0.813953 | 0.364016 |
| Obj_B | 0.714286 | 0.120000 | 0.023256 | 0.285847 |
| Obj_C | 0.047619 | 0.840000 | 0.162791 | 0.350137 |

**Best-Worst (BW) Method** is another subjective weighting method, introduced by Rezaei (2016) using two vectors of pair-wise comparisons to derive weights of objectives, which has been receiving much attention in recent years. It calculates the weight for each objective based on only the decision makers' subjective inputs, like AHP. The three steps in BW method are as follows.

**Step <1>:** Decision makers are to Identify the best objective (i.e., their most favored objective) and the worst objective (i.e., their least favored objective).

**Step <2>:** Decision makers are to decide the distances from the best objective that they identified in the prior step to all the remaining objectives, on the integer scale between 1 and 9. The Best-to-Others (BO) vector is then constructed as:

$$A_B = \{a_{B1}, a_{B2}, \dots, a_{Bn}\} \tag{15.9}$$

Here, $a_{Bj}$, within the range of [1, 9], is the distance from the best objective $B$ to objective $j$. The larger the $a_{Bj}$ value, the further the objective $j$ is from the best objective (i.e., in terms of priority). Note that $a_{BB}$ (i.e., best-to-best) is always 1, and $a_{Bj}$ need not be different for different objectives. Similarly, decision makers need to decide the distances of all objectives to the worst objective that they identified. The Others-to-Worst (OW) vector is then constructed as

$$A_W = \{a_{1W}, a_{2W}, \dots, a_{nW}\} \tag{15.10}$$

Here, $a_{jW}$, within the range of [1, 9], is the distance from objective $j$ to the worst objective $W$. $a_{WW}$ (i.e., worst-to-worst) is set as 1, and $a_{jW}$ need not be different for different objectives.

**Step <3>:** The final step of BW method is to formulate and solve the linear optimization problem, whose solution generates the weights for the objectives.

| | | | |
|---|---|---|---|
| *Minimize* | $\xi$ | | (15.11a) |
| *With respect to* | $w_1, w_2, \ldots, w_n$ | | (15.11b) |
| *Subject to* | $\sum_{j=1}^{n} w_j = 1$ | | (15.11c) |
| | $\lvert w_B - a_{Bj} w_j \rvert \leq \xi$ | $where\ j\ \in [1, n]$ | (15.11d) |
| | $\lvert w_j - a_{jW} w_W \rvert \leq \xi$ | $where\ j\ \in [1, n]$ | (15.11e) |

Here, $w_B$ and $w_W$ are the weight of the best and worst objectives, respectively.

## 15.4 Selected MCDM Methods

Our recent studies (Wang and Rangaiah, 2017; Wang et al., 2020) analyzed many MCDM methods for many mathematical problems and engineering applications. Considering simplicity of principle, user inputs required, ability to handle objective values of different magnitudes and popularity, we recommended the following MCDM methods: Technique for Order of Preference by Similarity to Ideal Solution (TOPSIS), Simple Additive Weighting (SAW), Gray Relational Analysis (GRA) and Multi-Attributive Border Approximation Area Comparison (MABAC). Recently, we proposed Preference Ranking On the Basis of Ideal-average Distance (PROBID) method (Wang et al., 2021). These 5 MCDM methods are described in the following sub-sections.

### 15.4.1 Gray Relational Analysis

Normalization of the objective matrix, identification of ideal reference network and computation of the gray relational coefficient (GRC) are the three key steps in the GRA method. The non-dominated solution (or alternative) with the highest GRC value is ranked as the top optimal solution and recommended. The GRA method has the unique feature of not requiring any user inputs, such as weights for the objectives. A numerical example illustrating the GRA method can be found in Song and Jamalipour (2005).

**Step <1>:** Normalize the objective matrix using the max-min normalization (Table 15.1), as in Song and Jamalipour (2005). Recall that normalized values of a cost objective by this normalization will be like those of a benefit objective.

**Step <2>:** Identify the ideal reference network having the best value for each of all the objectives.

$$F_j^+ = max_{i\in[1,m]} F_{ij} \qquad where\, j\, \in [1,n] \tag{15.12}$$

**Step <3>:** Compute GRC value for each non-dominated solution by the following equations.

$$\Delta I_{ij} = |F_j^+ - F_{ij}| \qquad where\, i\, \in [1,m]\, and\, j\, \in [1,n] \tag{15.13}$$

$$GRC_i = \frac{1}{m}\sum_{j=1}^{n}\frac{\Delta min+\Delta max}{\Delta I_{ij}+\Delta max} \qquad where\, i \in [1,m] \tag{15.14}$$

Here, $\Delta max = max_{i\in[1,m],j\in[1,n]}(\Delta I_{ij})$ and $\Delta min = min_{i\in[1,m],j\in[1,n]}(\Delta I_{ij})$. The solution having the highest GRC value is top ranked and recommended.

### 15.4.2 Multi-Attributive Border Approximation Area Comparison

Pamučar and Ćirović (2015) proposed MABAC method and illustrated it with a numerical example. The main idea behind it is to calculate the distance between each non-dominated solution and the boundary approximation area that is based on the product of weighted normalized values of each objective. The top ranked optimal solution is that with the greatest distance. The steps of MABAC are as follows.

**Step <1>:** Normalize the objective matrix using the max-min normalization (Table 15.1), as in Pamučar and Ćirović (2015). Recall that normalized values of a cost objective by this normalization will be like those of a benefit objective.

**Step <2>:** Apply weight to each objective; in this, $F_{ij}$ is offset by 1 to avoid any value becoming 0 in the next step.

$$v_{ij} = (1 + F_{ij}) \times w_j \quad i \in [1,m]\, and\, j \in [1,n] \tag{15.15}$$

**Step <3>:** Determine the boundary approximation area (by equation 15.16), and then calculate the distance of each non-dominated solution to the boundary approximation area (by equation 15.17).

$$b_j = (\prod_{i=1}^{m} v_{ij})^{1/m} \qquad j \in [1,n] \tag{15.16}$$

$$Q_i = \sum_{j=1}^{n}(v_{ij} - b_j) \quad i \in [1,m] \tag{15.17}$$

The solution/alternative with the greatest distance ($Q_i$) is top ranked and recommended.

### 15.4.3 Preference Ranking On the Basis of Ideal-average Distance

The primary benefit of PROBID method proposed by Wang et al. (2021) is that it provides a thorough coverage of the mean solution and many tiers of ideal solutions for ranking non-dominated solutions. See Wang et al. (2021) for detailed discussion about this method and a numerical example. The steps of PROBID are as follows:

**Step <1>:** Normalize the objective matrix using vector normalization (Table 15.1), as in Wang et al. (2021).

**Step <2>:** Apply weight to each objective.

$$v_{ij} = F_{ij} \times w_j \quad i \in [1, m] \text{ and } j \in [1, n] \tag{15.18}$$

**Step <3>:** Determine the most positive ideal solution, abbreviated as PIS ($A_{(1)}$), 2nd PIS ($A_{(2)}$), 3rd PIS ($A_{(3)}$), ..., and $m^{th}$ PIS ($A_{(m)}$) (i.e., negative ideal solution).

$$A_{(k)} = \{(Large(v_j, k) | j \in J), (Small(v_j, k) | j \in J')\} =$$
$$\{v_{(k)1}, v_{(k)2}, v_{(k)3}, \ldots, v_{(k)j}, \ldots, v_{(k)n}\} \tag{15.19}$$

Here, $k \in [1, m]$, $J$ is the set of maximization objectives from $[1, n]$, $J'$ is the set of minimization objectives from $[1, n]$, $Large(v_j, k)$ represents the $k^{th}$ largest value in the $j$th weighted normalized objective, $Small(v_j, k)$ represents the $k^{th}$ smallest value in the $j$th weighted normalized objective, and $v_{(k)j}$ is the $k^{th}$ best value (i.e., the $k^{th}$ largest value for a benefit objective or the $k^{th}$ smallest value for a cost objective) in the $j^{th}$ weighted normalized objective. The average value of each weighted normalized objective is calculated as:

$$\bar{v}_j = \frac{\sum_{k=1}^{m} v_{(k)j}}{m} \quad \text{for } j \in [1, n] \tag{15.20}$$

$$\bar{A} = \{\bar{v}_1, \bar{v}_2, \bar{v}_3, \ldots, \bar{v}_j, \ldots, \bar{v}_n\} \tag{15.21}$$

**Step <4>:** Compute the Euclidean distance of each non-dominated solution to each of the $m$ ideal solutions.

$$S_{i(k)} = \sqrt{\sum_{j=1}^{n} (v_{ij} - v_{(k)j})^2} \quad i, k \in [1, m] \tag{15.22}$$

The distance of each non-dominated solution to the average solution is computed as:

$$S_{i(avg)} = \sqrt{\sum_{j=1}^{n} (v_{ij} - \bar{v}_j)^2} \quad i \in [1, m] \tag{15.23}$$

**Step <5>:** Find the overall positive-ideal distance as well as the overall negative-ideal distance.

$$S_{i(pos-ideal)} = \begin{cases} \sum_{k=1}^{\frac{m+1}{2}} \frac{1}{k} S_{i(k)} \; i \in [1, m] \text{ when } m \text{ is an odd number} \\ \sum_{k=1}^{\frac{m}{2}} \frac{1}{k} S_{i(k)} \; i \in [1, m] \text{ when } m \text{ is an even number} \end{cases} \tag{15.24}$$

$$S_{i(neg-ideal)} = \begin{cases} \sum_{k=\frac{m+1}{2}}^{m} \frac{1}{m-k+1} S_{i(k)} \; i \in [1, m] \text{ when } m \text{ is an odd number} \\ \sum_{k=\frac{m}{2}+1}^{m} \frac{1}{m-k+1} S_{i(k)} \; i \in [1, m] \text{ when } m \text{ is an even number} \end{cases} \tag{15.25}$$

**Step <6>:** Compute the ratio of pos-ideal distance to neg-ideal distance ($R_i$) and then performance score ($P_i$) of each non-dominated solution.

$$R_i = \frac{S_{i(pos-ideal)}}{S_{i(neg-ideal)}} \quad i \in [1, m] \tag{15.26}$$

$$P_i = \frac{1}{1+R_i^2} + S_{i(avg)} \quad i \in [1, m] \tag{15.27}$$

The solution/alternative with the largest performance score is top ranked and recommended.

### 15.4.4 Simple Additive Weighting (SAW)

The SAW method, possibly the simplest MCDM methods, was first presented in Fishburn (1967) and MacCrimmon (1968). It begins by utilizing the max normalization to create the normalized objective matrix and applying weights. Then, objective values of each non-dominated solution (or alternative) are simply summed up. The top ranked solution is the one with the highest value in this summation and recommended. A numerical example about using the SAW method can be found in (Afshari et al., 2010).

**Step <1>:** Normalize the objective matrix using max normalization (Table 15.1), as in MacCrimmon (1968). Recall that normalized values of a cost objective by this normalization will be like those of a benefit objective. Owing to max normalization, one limitation of SAW method is that, in the original objective matrix, maximum value of a maximization (benefit) objective and any value of a minimization (cost) objective cannot be 0, in order to avoid division by zero.

**Step <2>:** Apply weight to each objective.

$$v_{ij} = F_{ij} \times w_j \qquad i \in [1,\ m] \text{ and } j \in [1,\ n] \tag{15.28}$$

**Step <3>:** Sum up the weighted normalized value for each non-dominated solution.

$$A_i = \sum_{j=1}^{n} v_{ij} \qquad i \in [1,\ m] \tag{15.29}$$

The solution/alternative with the largest value for the sum is top ranked and recommended.

### 15.4.5 Technique for Order of Preference by Similarity to Ideal Solution

Hwang and Yoon (1981) created TOPSIS method and illustrated it using a numerical example. The top ranked non-dominated solution chosen in this method is the solution that is furthest to the negative ideal solution while also being the closest to the positive ideal solution. The best value of each objective, that is, the maximal value of the maximization objective and the minimal value of the minimization objective, makes up the ideal positive solution. On the other hand, the worst value of each objective, that is, the minimal value of the maximization objective and the maximal value of the minimization objective, constitute the negative ideal solution. The steps in TOPSIS method are as follows.

**Step <1>:** Normalize the objective matrix using vector normalization (Table 15.1), as in Hwang and Yoon (1981).

**Step <2>:** Apply weight to each objective.

$$v_{ij} = F_{ij} \times w_j \qquad i \in [1,\ m] \text{ and } j \in [1,\ n] \tag{15.30}$$

**Step <3>:** Determine the positive ideal solution $A^+$ and negative ideal solution $A^-$ as follows

$$A^+ = \{(max_{i \in [1,m]}(v_{ij})|j \in J), (min_{i \in [1,m]}(v_{ij})|j \in J')\}$$
$$= \{v_1^+, v_2^+, v_3^+, \ldots, v_j^+, \ldots, v_n^+\} \tag{15.31}$$

$$A^- = \{(min_{i\in[1,m]}(v_{ij})|j \in J), (max_{i\in[1,m]}(v_{ij})|j \in J')\}$$
$$= \{v_1^-, v_2^-, v_3^-, \ldots, v_j^-, \ldots, v_n^-\} \tag{15.32}$$

Here, $J$ is the set of maximization objectives from $[1, n]$ and $J'$ is the set of minimization objectives from $[1, n]$.

**Step <4>:** Compute the Euclidean distance of each non-dominated solution to the positive ideal and negative ideal solutions, respectively, as follows:

$$S_{i+} = \sqrt{\sum_{j=1}^{n}(v_{ij} - v_j^+)^2} \quad i \in [1, m] \tag{15.33}$$

$$S_{i-} = \sqrt{\sum_{j=1}^{n}(v_{ij} - v_j^-)^2} \quad i \in [1, m] \tag{15.34}$$

**Step <5>:** Quantify the closeness ($C_i$) of each non-dominated solution.

$$C_i = \frac{S_{i-}}{S_{i-} + S_{i+}} \quad i \in [1, m] \tag{15.35}$$

The solution with the largest $C_i$ is top ranked and recommended.

### 15.5 Microsoft Excel Program for MCDM

Calculations for the 4 normalization methods, 4 weighting methods and 5 MCDM methods described in the above sections can be performed using a calculator or a spreadsheet. However, a validated and easy-to-use program for them will be very useful for correct and quick computations. Hence, the 4 normalization, 4 weighting and 5 MCDM methods are implemented in a Microsoft Excel program: EMCDM445, whose Overview worksheet is in Figure 15.3. Interested readers can obtain a copy of it by sending an email request to chegpr@nus.edu.sg and wangzhiyuan@u.nus.edu. A more extensive Microsoft Excel program with 8 weighting methods and 15 MCDM methods (Wang et al., 2020; 2021) is also available via an email request. Both the programs have a friendly user interface and are easy to use for MCDM.

| | A | B | C |
|---|---|---|---|
| 1 | | \multicolumn{2}{l|}{**Methods for Ranking Pareto-Optimal (Non-Dominated) Solutions to Select One Optimal Solution**} |
| 2 | | \multicolumn{2}{l|}{Department of Chemical and Biomolecular Engineering, National University of Singapore} |
| 3 | | \multicolumn{2}{l|}{(https://blog.nus.edu.sg/rangaiah/)} |
| 4 | | \multicolumn{2}{l|}{For more details, see Wang Z., Nabavi S.R. and Rangaiah G.P., "Selected Multi-Criteria Decision Making Methods and their Applications to Product and System Designs" in the book by A.J. Kulkarni (editor), 'Optimization Methods for Product and System Design', Springer (2022).} |
| 5 | | \multicolumn{2}{l|}{Hope you can use this program in your work. Please cite the above reference in your publications.} |
| 6 | | | |
| 7 | | Worksheet | Description |
| 8 | | Overview | 1. The overview serves to provide brief description and instructions for using this program.<br>2.. Carefully read and follow them. |
| 9 | | Results From MOO | 1. This worksheet is for the user to provide Pareto-optimal (i.e., non-dominated) solutions for his/her application.<br>2. This program requires non-dominated solutions, which should be copied to this worksheet.<br>3. Number of non-dominated solutions can be few (e.g., 4 or 9) and many (e.g., 78, 105 or 164).<br>4. There are no limits on number of objectives, decision variables, constraints and non-dominated/optimal solutions.<br>5. Data on decision variables and constraints are optional for using this program.<br>6. Names of Decision Variables (X1, X2, X3 etc.), Objectives (F1, F2 etc.) and Constraints (C1, C2, C3 etc.), if available, should be pasted in row 3 of worksheet "Results From MOO".<br>7. All values should be pasted in rows starting from row 5. Closely and strictly follow the format in the current sheet "Results from MOO". |
| 10 | | Main Interface | 1. In this main interface of the program, user needs to provide essential information such as Number of Objectives, Decision Variables, Constraints etc.<br>2. Number of Decision Variables and Constraints should be given as 0 (zero) if user does not have those data.<br>3. User can choose any weighting method or ranking/election method; s/he can also enter their own inputs (i.e., without using any of the weighting methods).<br>4. If user wishes to try all rankin/selection methods, click 'Run all Methods' and then click 'Show all Charts' to have results of all methods in the form of plots for comparison.<br>5. All created worksheets can be deleted by clicking 'Delete Worksheets of All Methods'. Use this as and when required before trying a different weighting and/or ranking/selection method. |
| 11 | | Results of Each Method | Results of each method or all methods chosen by the user from the Main Interface are given on newly created worksheets (e.g, TOPSIS Results). Scroll down and/or to right to see the plots. |
| 12 | | All Charts | This worksheet will be created if 'Show All Charts' is clicked. |
| 13 | | | |

Figure 15.3: Overview worksheet of the EMCDM445 Program

The EMCDM445 has the following features.

- Objective matrix data are provided by the user in the worksheet "Results from MOO" (Figure 15.4).

- It can compute weights of objectives by one of the objective or subjective weighting methods, as per user's choice. Alternatively, user can give her/his preferred weights. Further, the program has the option to combine the weights found by a weighting method with those given by the user, via the formula: combined weight of jth objective $= \frac{W_{uj}W_{wj}}{\sum_{j=1}^{n} W_{uj}W_{wj}}$, where $W_{uj}$ is the user-provided weight and $W_{wj}$ is the weight found by the chosen weighting method, both for the jth objective.

- The user can choose to run any one of the 5 MCDM methods by clicking it. Alternatively, she/he can run all 5 MCDM methods by clicking "Run all Methods" (at the bottom left in Figure 15.5).

- The default normalization for each MCDM method is as per its original reference; see the description in Section 15.3. However, the program allows the user to change the normalization method (to another that is feasible with the MCDM method) or choose not to use normalization (i.e., apply MCDM method on the original dataset).

- The "Show All Charts" button (at the bottom left in Figure 15.5) is to consolidate and present all the charts/results of 5 MCDM methods that were run, in one worksheet. To succinctly present objective values of different magnitude in one plot, fractional objective values are calculated using: $\text{Fractional jth objective} = \frac{F_j - F_{j,min}}{F_{j,max} - F_{j,min}}$.

- The program allows the user to decide whether to plot or not.

- The "Delete All Results" button (at the bottom left in Figure 15.5) is to safely delete all the MCDM worksheets containing the charts/results of previous runs. Use this button before a new trial or run.

| | A | B | C | D | E | F | G | H |
|---|---|---|---|---|---|---|---|---|
| 1 | | | | | | | | |
| 2 | | Objective Matrix (Dataset) for Machine Tool Selection | | | | | | |
| 3 | Objective Names → | CC (Min) | SS (Max) | TC (Max) | TX (Min) | TZ (Min) | MD (Max) | ML (Max) |
| 4 | Alternatives↓ | | Leave this row 4 as blank. | | | | | |
| 5 | 1 | 1200000 | 5590 | 8 | 24 | 24 | 205 | 350 |
| 6 | 2 | 1550000 | 3465 | 8 | 20 | 20 | 280 | 520 |
| 7 | 3 | 1400000 | 5950 | 12 | 15 | 20 | 250 | 469 |
| 8 | 4 | 1100000 | 5940 | 12 | 12 | 15 | 230 | 600 |
| 9 | 5 | 1200000 | 5940 | 12 | 12 | 16 | 150 | 330 |
| 10 | 6 | 1500000 | 3465 | 12 | 6 | 12 | 260 | 420 |
| 11 | 7 | 2600000 | 3960 | 12 | 12 | 16 | 300 | 625 |
| 12 | 8 | 1320000 | 4950 | 12 | 24 | 30 | 240 | 340 |
| 13 | 9 | 1180000 | 4480 | 8 | 24 | 24 | 250 | 330 |
| 14 | 10 | 1550000 | 3950 | 12 | 15 | 20 | 280 | 460 |
| 15 | 11 | 1600000 | 3450 | 12 | 15 | 20 | 280 | 460 |
| 16 | 12 | 1200000 | 3465 | 8 | 20 | 24 | 264 | 400 |
| 17 | 13 | 1350000 | 2970 | 8 | 20 | 24 | 264 | 400 |
| 18 | 14 | 1400000 | 2970 | 12 | 24 | 30 | 300 | 600 |
| 19 | 15 | 1350000 | 3465 | 12 | 30 | 30 | 264 | 350 |
| 20 | 16 | 1450000 | 2970 | 12 | 20 | 24 | 300 | 400 |
| 21 | 17 | 1520000 | 2475 | 12 | 20 | 24 | 300 | 400 |
| 22 | 18 | 1376000 | 4752 | 12 | 20 | 24 | 235 | 350 |
| 23 | 19 | 1440000 | 4752 | 12 | 20 | 24 | 235 | 600 |
| 24 | 20 | 1824000 | 3790 | 10 | 12 | 20 | 300 | 530 |
| 25 | 21 | 1920000 | 3790 | 10 | 12 | 20 | 300 | 1030 |
| 26 | | | | | | | | |

Figure 15.4: Worksheet for providing the objective matrix data to the EMCDM445 Program

Before using the program for MCDM, study the important points in the Overview worksheet (Figure 15.3). This is essential for correct use of the program. As shown in Figure 15.5, there are 5 steps for performing MCDM.

**Step <1>:** Provide the objective matrix (i.e., values of all objectives for the various alternatives) in the worksheet: "Results from MOO". Figure 15.4 presents this worksheet with values for 7 objectives and 21 alternatives, for the machine tool selection problem. Alternatively, Pareto-optimal solutions found by multi-objective optimization can be provided in the "Results from MOO" worksheet; these can include values of decision variables and constraints besides the objective values. In either case, values of the first alternative/solution must be in row 5; this is important. Names of objectives can be given in columns B, C, … of row 3 (Figure 15.4).

**Step <2>:** Enter number of objectives (must be 2 or more), number of decision variables (zero if none), number of constraints (zero if none) and number of alternatives/solutions (must be 2 or more) in cells D5 to D8 (Figure 15.5).

**Step <3>:** Choose the type (either Max or Min) of each objective.

**Step <4>:** Weight can be given by the user or found by one of the 4 methods on the right side in Figure 15.5.

**Step <5>:** Perform MCDM by choosing one or all 5 MCDM methods on the left side in Figure 15.5.

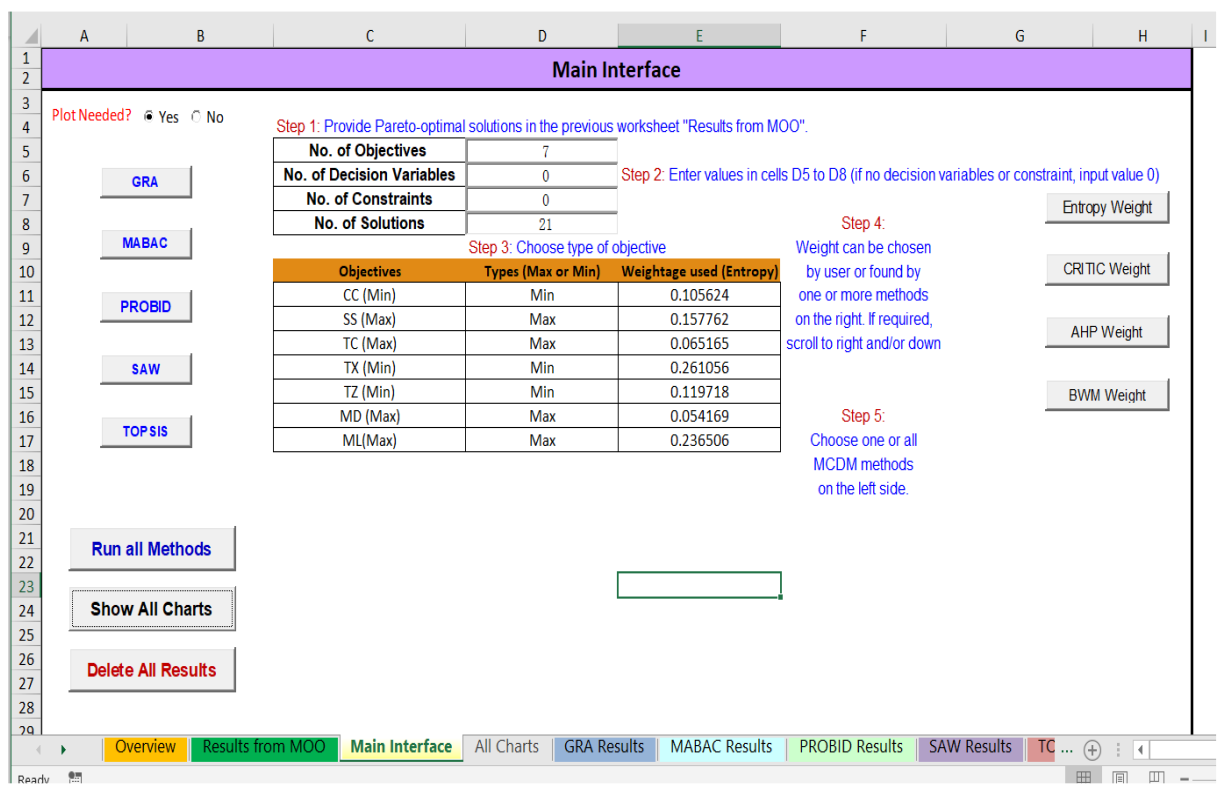

Figure 15.5: Main Interface of the EMCDM445 program

Results of each MCDM method are presented in a separate worksheet, as can be seen by different worksheets in the bottom of Figure 15.5. If "Show All Charts" is executed, selected solution/alternative by the 5 methods are presented in the "All Charts" worksheet, which includes objective values of the selected alternatives in the form of a bar chart, a radar chart and a table (Figure 15.6) as well as 2D plots (one for each MCDM method, not shown in Figure 15.6) of the first and last objectives. These enable the user to assess, both qualitatively and quantitatively, the alternatives (Pareto-optimal solutions) recommended by different MCDM methods.

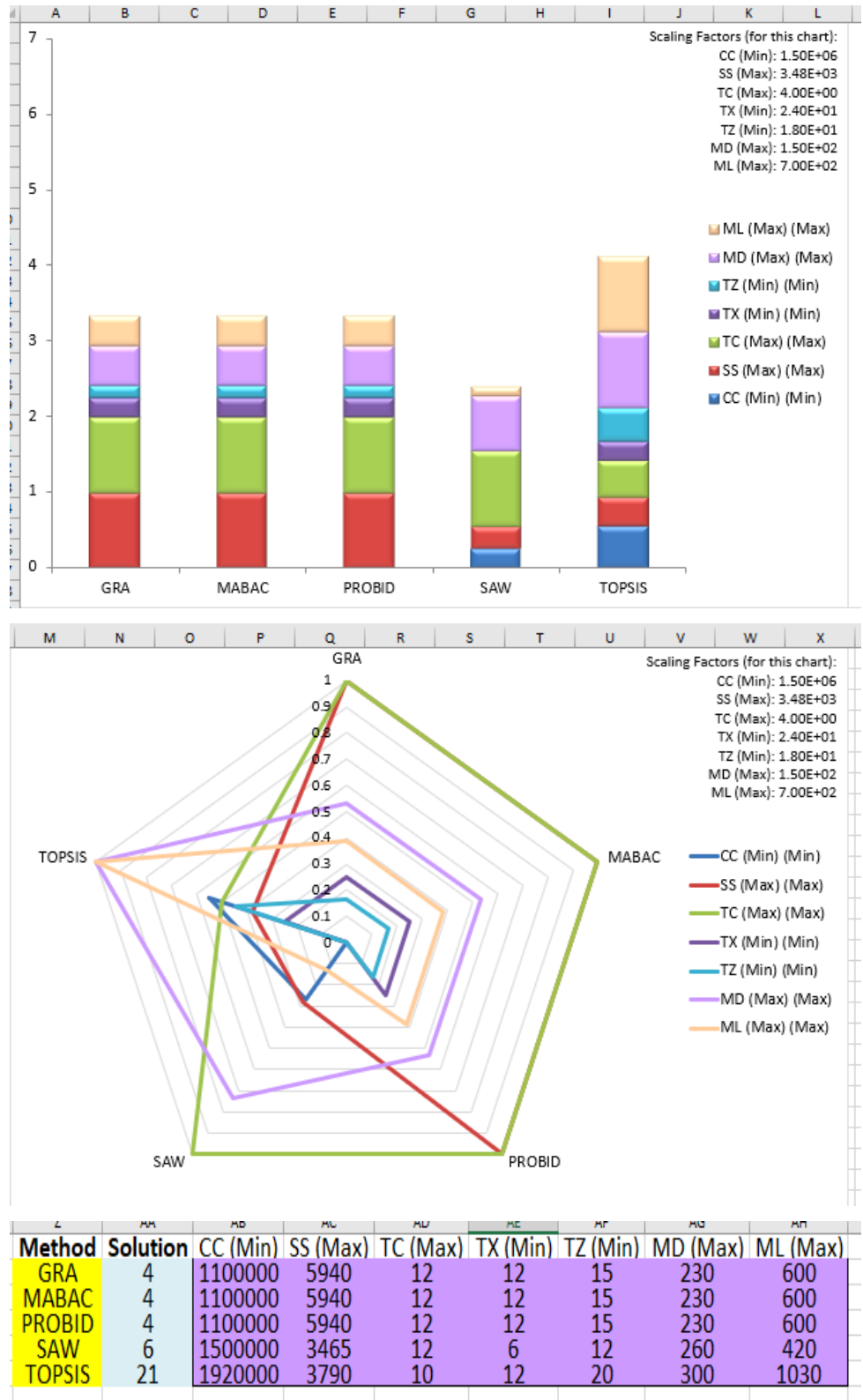

Figure 15.6: Bar Chart, Radar Chart and List of Selected Solutions by GRA, MABAC, PROBID, SAW and TOPSIS, in the "All Charts" worksheet of the EMCDM445 program

## 15.6 Application of MCDM Methods to Product and System Designs

For illustrating the 4 weighting methods and 5 MCDM methods as well as EMCDM445 program, 4 applications from product and system designs are carefully chosen. Brief details of these applications are given in Table 15.4 whereas their objective matrices (datasets) are in the Appendix. As can be seen from Table 15.4, the selected 4 applications are diverse in the area/domain of the application, number of objectives from 2 to 10, and number of alternatives from 6 to 74. Moreover, the objective matrix for fruit supply chain and VOC recovery applications was obtained in the cited reference, by solving the related multi-objective optimization problem. On the other hand, the objective matrix for machine tool selection and non-traditional machining (NTM) process selection were probably compiled from the available alternatives and their values for the performance criteria (objectives). However, in all these applications, ranking and selection of one alternative (Pareto-optimal solution) requires MCDM. Here, each application was analyzed using 5 MCDM methods; for conciseness, only 2 of the 4 weighting methods were used in each application. The results of these trials are sufficient to illustrate the weighting methods and MCDM methods described above.

Table 15.4 Brief Details of Selected MCDM Applications from Product and System Design; See Tables 15.5 to 15.8 for the objectives/criteria in each application.

| Application: Brief Details | No. of Objectives | No. of Alternatives | Reference(s) |
|---|---|---|---|
| **Fruit Supply Chain**: This supply hain was optimized for simultaneous minimization of total cost and maximization of satisfying demand (quantified as responsiveness). | 2 | 16 from the Pareto-optimal front | Cheraghalipour et al. (2018) |
| **VOC Recovery**: Design of a process system for recovery of volatile organic compounds (VOC) considering both economic and environmental objectives. | 5 | 74 from the Pareto-optimal front | Lee and Rangaiah (2009) |
| **Machine Tool Selection**: This application is on selecting one from many alternatives for a computerized numerical control machine. | 7 | 16 non-dominated alternatives | Sun (2002); Chakraborty (2011) |

| | | | | | |
|---|---|---|---|---|---|
| **Non-Traditional Machining (NTM) Process Selection:** This application is on the selection of one from various machining processes for the treatment of advanced engineering materials | 10 | 8 non-dominated alternatives | Chakraborty (2011) | | |

Results from the application of 5 MCDM methods with the chosen weighting methods are presented in Tables 15.5 to 15.8. The recommended alternative/solution by GRA is unaffected by the weighting method (as it does not require any weights). Recommended alternative by an MCDM method may be same or different from that by GRA because of different weights and/or MCDM method. The differences, if any, between the alternative recommended by an MCDM method and that by GRA are quantified by the average of absolute fractional differences (AAFD) between the values of criteria in the two recommended alternatives; it is given by:

$$\text{AAFD} = \frac{1}{n}\sum_{1}^{n}\frac{|C_{i,MCD}-C_{i,GRA}|}{|C_{i,MCD}+C_{i,GRA}|/2} \qquad (15.36)$$

Here, $n$ is the number of criteria/objectives, $C_{i,MCD}$ is the value of the i$^{th}$ criterion in the alternative recommended by an MCD method and $C_{i,GRA}$ is the value of the i$^{th}$ criterion in the alternative recommended by GRA. The value of AAFD indicates the closeness of the alternative recommended by an MCD method with that by GRA; zero or small value of AAFD means the two recommended alternatives are same or comparable.

Table 15.5  Recommended Alternatives by 5 MCDM Methods with 2 Different Weighting Methods: Fruit Supply Chain Dataset

| Objective | Weight | GRA | MABAC | PROBID | SAW | TOPSIS |
|---|---|---|---|---|---|---|
| | *Using Weights by the Entropy Method* | | | | | |
| Min. TC | 0.886324 | 5356513 | 5272345 | 5324449 | 5272345 | 5272345 |
| Max. RS | 0.113676 | 0.776971 | 0.745228 | 0.764647 | 0.745228 | 0.745228 |
| *AAFD* | | *Nil* | 0.028772 | 0.010996 | 0.028772 | 0.028772 |
| | *Using Weights by AHP Method** | | | | | |
| Min. TC | 0.125 | 5356513 | 6755311 | 5941683 | 5941683 | 5941683 |
| Max. RS | 0.875 | 0.776971 | 0.822531 | 0.812448 | 0.812448 | 0.812448 |
| *AAFD* | | *Nil* | *0.143974* | *0.074114* | *0.074114* | *0.074114* |

Notes: TC – Total Cost; RS – Responsiveness.

* In AHP method, favor level of C2 (RS) over C1 (TC) was chosen as 7.

For the fruit supply chain dataset, recommended alternatives by 5 MCDM methods with entropy and AHP weights are presented in Table 15.5. As evident from this table, weights of objectives (column 2) by the entropy and AHP methods are very different. One reason for this is the chosen favor level of 7 for RS over TC. Weights by AHP method would be equal if the favor level of RS over TC is zero. Recall that AHP method is a subjective method, and its weights depend on the favor level selected by the user.

In the case of weights by the entropy method, the recommended alternatives for fruit supply chain application by 5 MCDM methods are comparable (Table 15.5). On the other hand, use of weights by AHP method leads to recommendation of a different alternative by MABAC, PROBID, SAW and TOPSIS compared to that by GRA. Interestingly, PROBID, SAW and TOPSIS recommend the same alternative. This may not always occur for two reasons: weights by AHP method depend on user's inputs and different application (i.e., objective matrix). Results in Table 15.5 indicate that the recommended alternative by a MCDM method can be different depending on the weights.

For VOC recovery process design, recommended alternatives by the 5 MCDM methods with CRITIC and BW methods weights are given in Table 15.6. Weights of two objectives (namely, NPW and PCOP) are comparable by both CRITIC and BW methods whereas those of the remaining objectives (namely, HTP, ETP and ATMP) are different. Recall that BW method is based in user's inputs given in the footnote of Table 15.6. In the case of CRITIC weights, the 5 MCDM methods recommend different alternatives; however, PROBID and TOPSIS recommend the same alternative. Using weights by BW method, the recommended alternative by MABAC is the same as that of GRA. PROBID and TOPSIS select the same alternative, and top-ranked alternative by SAW is the same as that using CRITIC weights. These observations are likely to be different if the user gives different inputs to BW method.

Table 15.6   Recommended Alternatives by 5 MCDM Methods with 2 Different Weighting Methods: VOC Recovery Process Design

| Objective | Weight | GRA | MABAC | PROBID | SAW | TOPSIS |
|---|---|---|---|---|---|---|
| | *Using Weights by the CRITIC Method* | | | | | |
| Max. NPW | 0.176857 | 5698102.8 | 4115691.5 | 3557449.5 | 3528699 | 3557449.5 |
| Min. HTP | 0.222063 | 1.732E-07 | 1.388E-07 | 1.584E-07 | 1.552E-07 | 1.584E-07 |
| Min. ETP | 0.164652 | 3.024E-05 | 3.148E-05 | 4.276E-05 | 4.31E-05 | 4.276E-05 |
| Min. ATMP | 0.163876 | 0.2026962 | 0.2111912 | 0.2869329 | 0.2892404 | 0.2869329 |
| Min. PCOP | 0.272552 | 13.631483 | 7.4598343 | 0.5780334 | 0.5259401 | 0.5780334 |
| *AAFD* | | *Nil* | *0.241892* | *0.615242* | *0.626762* | *0.615242* |
| | *Using Weights by BW Method\** | | | | | |
| *Max. NPW* | 0.188679 | *5698102.8* | *5698102.8* | 3091184.1 | 3528699 | 3091184.1 |
| *Min. HTP* | 0.05283 | *1.732E-07* | *1.732E-07* | 1.531E-07 | 1.552E-07 | 1.531E-07 |
| *Min. ETP* | 0.362264 | *3.024E-05* | *3.024E-05* | 3.902E-05 | 4.31E-05 | 3.902E-05 |
| *Min. ATMP* | 0.113208 | *0.2026962* | *0.2026962* | 0.2617941 | 0.2892404 | 0.2617941 |
| *Min. PCOP* | 0.283019 | *13.631483* | *13.631483* | 1.6027706 | 0.5259401 | 1.6027706 |
| *AAFD* | | *Nil* | *0.0* | 0.560661 | 0.626689 | 0.560661 |

*Notes: NPT – Net Present Worth; HTP – Human Toxicity Potential; ETP – Ecotoxicity Potential; Atmospheric Potential; PCOP – Photochemical Oxidation Potential*

\* Inputs given for BW method are as follows: Best Objective is ETP (C3), Worst Objective is HTP (C2), Best-to-Others (BO) vector (equation 15.9) is (3, 4, 1, 5, 2) and Worst-to-Others (WO) vector (equation 15.10) is (4, 1, 3, 6, 2).

For the machine tool selection, Table 15.7 summarizes the recommended alternatives by 5 MCDM methods with weights by the entropy and BW methods, which give different weights for the objectives in this application. Using the entropy method, GRA, MABAC and PROBID recommend the same alternative whereas SAW and TOPSIS recommend a different alternative. In the case of weights by BW method, all 5 MCDM methods lead to the same alternative; however, this may not occur with other preferences of the user provided to BW method. Recall that alternatives recommended by 5 MCDM methods with weights BM method are not the same for the VOC recovery application in Table 15.6.

Table 15.7   Recommended Alternatives by 5 MCDM Methods with 2 Different Weighting Methods: Machine Tool Selection

| Objective | Weight | GRA | MABAC | PROBID | SAW | TOPSIS |
|---|---|---|---|---|---|---|
| | *Using Weights by the Entropy Method* | | | | | |
| Min. CC | 0.120161 | 1100000 | 1100000 | 1100000 | 1500000 | 1920000 |
| Max. SS | 0.114203 | 5940 | 5940 | 5940 | 3465 | 3790 |
| Max. TC | 0.055164 | 12 | 12 | 12 | 12 | 10 |
| Min. TX | 0.298785 | 12 | 12 | 12 | 6 | 12 |
| Min. TZ | 0.137903 | 15 | 15 | 15 | 12 | 20 |
| Max. MD | 0.023069 | 230 | 230 | 230 | 260 | 300 |
| Max. ML | 0.250715 | 600 | 600 | 600 | 420 | 1030 |
| *AAFD* | | *Nil* | 0 | 0 | 0.314041 | 0.32061 |

|  | Using Weights by BW Method* | | | | | |
|---|---|---|---|---|---|---|
| Min. CC | 0.336328 | 1100000 | 1100000 | 1100000 | 1100000 | 1100000 |
| Max. SS | 0.164063 | 5940 | 5940 | 5940 | 5940 | 5940 |
| Max. TC | 0.098438 | 12 | 12 | 12 | 12 | 12 |
| Min. TX | 0.030078 | 12 | 12 | 12 | 12 | 12 |
| Min. TZ | 0.246094 | 15 | 15 | 15 | 15 | 15 |
| Max. MD | 0.054688 | 230 | 230 | 230 | 230 | 230 |
| Max. ML | 0.070313 | 600 | 600 | 600 | 600 | 600 |
| AAFD |  | Nil | 0.0 | 0.0 | 0.0 | 0.0 |

Notes: CC – Capital Cost; SS – Spindle Speed Range; TC – Total Capacity; TX – Rapid Traverse Rate of X-axis; TZ – Rapid Traverse Rate of Z-Axis; MD – Maximum Machining Diameter; ML – Maximum Machining Length

* Inputs given for BW method are as follows: Best Objective is CC (C1), Worst Objective is TX (C4), Best-to-Others (BO) vector (equation 15.9) is (1, 3, 5, 6, 2, 9, 7) and Worst-to-Others (WO) vector (equation 15.10) is (6, 2, 8, 1, 4, 7, 6).

For NTM process selection, recommended alternatives by 5 MCDM methods in conjunction with 2 different methods for weights are presented in Table 15.8 whereas favor level given for objective functions for AHP method (like in Figure 15.2) are given in Table 15.9. In this application with 10 objectives, number of favor levels required for AHP method is 10×9/2 = 45. For NTM process selection, weights by CRITIC method are different from those by AHP method for some objectives, and they are comparable for other objectives. If weights by the CRITIC method are used, there are 3 recommended alternatives: one is given by both GRA and MABAC, another is given by PROBID and TOPSIS, and yet another is given by SAW. Using AHP weights, 4 of 5 MCDM methods give the same recommended solution. However, SAW gives a different alternative that is same as that recommended by it with CRITIC weights. Similar to BW method, the user may get different recommended alternatives with different inputs for AHP method.

Table 15.8  Recommended Alternatives by 5 MCDM Methods with 2 Different Weighting Methods: NTM Process Selection

| Objective | Weight | GRA | MABAC | PROBID | SAW | TOPSIS |
|---|---|---|---|---|---|---|
|  | Using Weights by the CRITIC Method | | | | | |
| Min.TSF | 0.076097 | 2 | 2 | 1 | 2.5 | 1 |
| Min. PR | 0.07372 | 1.4 | 1.4 | 10 | 0.2 | 10 |
| Max. MRR | 0.156419 | 0.1 | 0.1 | 500 | 1.6 | 500 |
| Min. C | 0.068471 | 3 | 3 | 2 | 4 | 2 |
| Max. E | 0.065112 | 5 | 5 | 4 | 5 | 4 |
| Min.TF | 0.104739 | 2 | 2 | 2 | 2 | 2 |
| Min.TC | 0.134625 | 1 | 1 | 3 | 1 | 3 |
| Max. S | 0.123594 | 3 | 3 | 1 | 3 | 1 |
| Max. M | 0.09471 | 5 | 5 | 5 | 5 | 5 |
| Max. F | 0.102514 | 5 | 5 | 5 | 5 | 5 |
| *AAFD* | | *Nil* | *0.0* | *0.679686* | *0.377264* | *0.679686* |
| | *Using Weights by AHP Method* | | | | | |
| Min. TSF | 0.07352 | 2 | 2 | 2 | 2.5 | 2 |
| Min. PR | 0.07185 | 1.4 | 1.4 | 1.4 | 0.2 | 1.4 |
| Max. MRR | 0.071435 | 0.1 | 0.1 | 0.1 | 1.6 | 0.1 |
| Min. C | 0.119381 | 3 | 3 | 3 | 4 | 3 |
| Max. E | 0.119695 | 5 | 5 | 5 | 5 | 5 |
| Min.TF | 0.060787 | 2 | 2 | 2 | 2 | 2 |
| Min.TC | 0.126612 | 1 | 1 | 1 | 1 | 1 |
| Max. S | 0.11549 | 3 | 3 | 3 | 3 | 3 |
| Max. M | 0.098725 | 5 | 5 | 5 | 5 | 5 |
| Max. F | 0.142505 | 5 | 5 | 5 | 5 | 5 |
| *AAFD* | | *Nil* | *0.0* | *0.0* | *0.377264* | *0.0* |

Notes: TSF – Tolerance and Surface Finish; PR – Power Requirement; MRR – Material Removal Rate; C – Cost; E – Efficiency, TF – Tooling and Fixtures; TC – Tool Consumption; S – Safety; M – Work Material, F – Shape Feature

Table 15.9. AHP parameters (favor level of objective functions) used for NTM dataset presented in Table 15.8.

| | C1 (TSF) | C2 (PR) | C3 (MRR) | C4 (C) | C5 (E) | C6 (TF) | C7 (TC) | C8 (S) | C9 (M) | C10 (IF) |
|---|---|---|---|---|---|---|---|---|---|---|
| C1 (TSF | | | | | | | | | | |
| C2 (PR) | 9 | | | | | | | | | |
| C3(MRR) | 7 | 5 | | | | | | | | |
| C4(C) | -3 | -7 | 9 | | | | | | | |
| C5(E) | 5 | 3 | -5 | -7 | | | | | | |
| C6(TF) | 0 | 9 | -3 | 3 | -9 | | | | | |
| C7(TC) | -7 | 0 | 7 | 0 | -5 | 5 | | | | |
| C8(S) | -3 | 9 | 0 | 5 | 7 | 9 | -7 | | | |
| C9(M) | 5 | 7 | 9 | -5 | -7 | 0 | 3 | 3 | | |
| C10(IF) | 9 | 3 | 5 | -3 | 5 | 0 | -5 | 9 | 5 | |

Results for 4 applications in Tables 15.5 to 15.8 by 5 MCDM methods and different weighting methods clearly show that weights can affect the recommended alternative by a MCDM method (except for GRA that does not require weights) and the recommended alternatives by 5 MCDM methods can be different or similar. Given these, the user can choose a particular weighting method and MCDM method, and then perform MCDM for the application. Here, the selection of a weighting method and MCDM method is subjective and not easy. Alternatively, the user can try several weighting methods and MCDM methods for the application and then review the recommended alternatives for choosing one of them. The recommended alternatives will not be many, and so their review is easier compared to reviewing all alternatives in the objective matrix. For this approach, we suggest the entropy and CRITIC methods (since they do not require inputs from the user) and 5 MCDM methods described in this chapter.

## 15.7 Summary

This chapter described the MCDM procedure and its many details including 4 normalization methods, 4 weighting methods and 5 MCDM methods, an Excel program for these methods and their use for 4 diverse applications. Learning points of this chapter are as follows.

- MCDM is required for choosing one of the alternatives (which may be the available choices, or Pareto-optimal or non-dominated solutions found by solving a multi-objective optimization problem).
- MCDM generally involves several steps: normalization of the objective matrix, use of a weighting method and MCDM method for ranking the alternatives.
- There are 4 normalization methods (Table 15.1), many weighting methods and many MCDM methods.
- Principles and equations of the entropy, CRITIC, AHP and BW weighting methods are presented in Section 15.3 whereas those of 5 MCDM methods (namely, GRA, MABAC, PROBID, SAW and TOPSIS) in Section 15.4.

- MS Excel-based program: EMCDM445 can be employed for MCDM. It has 4 normalization methods, 4 weighting methods and 5 MCDM methods; the user can choose any one of them for her/his application.
- EMCDM445 was employed for 4 diverse objective matrices (datasets) in Table 15.4. The results on weights of objectives and recommended alternatives are in Tables 15.5 to 15.8.
- The recommended solution for an application depends on the chosen weighting method as well as on MCDM method employed.
- For an application, it is not easy to choose one of the available weighting methods and one of many MCDM methods. Hence, use several MCDM methods in conjunction with a few weighting methods, and then analyze the recommended alternatives for choosing one of them. This analysis is easier compared to analyzing all alternatives in the objective matrix.
- Further studies are required for establishing the appropriate weighting and MCDM methods for an application.

**Appendix**

Table A.1  Fruit Supply Chain Dataset

| Alternatives | Cost (Min) | Responsiveness (Max) |
|---|---|---|
| 1 | 6755310.62 | 0.82253112 |
| 2 | 5272344.69 | 0.745228216 |
| 3 | 5356513.03 | 0.776970954 |
| 4 | 6290380.76 | 0.814688797 |
| 5 | 5941683.37 | 0.812448133 |
| 6 | 5789378.76 | 0.800497925 |
| 7 | 5536873.75 | 0.788174274 |
| 8 | 6611022.04 | 0.816182573 |
| 9 | 5304408.82 | 0.752697095 |
| 10 | 6077955.91 | 0.814315353 |
| 11 | 5324448.90 | 0.764647303 |
| 12 | 5689178.36 | 0.791908714 |
| 13 | 5336472.95 | 0.768008299 |
| 14 | 6663126.25 | 0.819170124 |
| 15 | 5713226.45 | 0.799004149 |
| 16 | 5601002.00 | 0.788921162 |

Table A.2  VOC Recovery Dataset

| Alternatives | NPW (Max) | HTP (Min) | ETP (Min) | ATMP (Min) | PCOP (Min) |
|---|---|---|---|---|---|
| 1 | 6281790.929 | 7.30988E-07 | 3.00506E-05 | 0.19937244 | 21.03959542 |
| 2 | 6229166.687 | 6.84412E-07 | 3.00814E-05 | 0.199751359 | 20.41826937 |
| 3 | 5922622.11 | 6.22614E-07 | 3.05126E-05 | 0.202878371 | 19.50919131 |
| 4 | 5750640.75 | 4.65311E-07 | 3.01689E-05 | 0.201146474 | 17.51728608 |
| 5 | 5698102.756 | 1.73162E-07 | 3.02393E-05 | 0.202696164 | 13.63148329 |
| 6 | 5693883.702 | 7.10462E-07 | 3.00838E-05 | 0.19967136 | 20.75485036 |
| 7 | 5659683.09 | 7.21214E-07 | 3.00848E-05 | 0.199638818 | 20.90722464 |
| 8 | 5597224.184 | 7.15728E-07 | 3.00829E-05 | 0.199645904 | 20.81855419 |
| 9 | 5566530.954 | 6.5288E-07 | 3.01385E-05 | 0.200251419 | 19.73823207 |
| 10 | 5515474.577 | 2.81644E-07 | 3.02528E-05 | 0.202387141 | 14.96397377 |
| 11 | 5497221.833 | 5.70282E-07 | 3.01166E-05 | 0.200408165 | 18.9177645 |
| 12 | 5469174.673 | 5.39636E-07 | 3.01247E-05 | 0.200575944 | 18.51884853 |
| 13 | 5378025.227 | 4.60758E-07 | 3.01775E-05 | 0.201221118 | 17.45760844 |
| 14 | 5351461.97 | 1.90471E-07 | 3.03431E-05 | 0.203330524 | 13.61091338 |
| 15 | 5256490.899 | 3.79832E-07 | 3.01838E-05 | 0.201561748 | 16.41466406 |
| 16 | 5094515.829 | 1.4562E-07 | 3.07972E-05 | 0.206548796 | 11.61647499 |
| 17 | 4996388.001 | 1.75383E-07 | 3.05412E-05 | 0.204718042 | 12.16176942 |
| 18 | 4988224.302 | 1.69023E-07 | 3.76726E-05 | 0.252694105 | 8.458640946 |
| 19 | 4812629.31 | 4.0948E-07 | 3.01596E-05 | 0.201289875 | 16.77191543 |
| 20 | 4788595.762 | 4.73031E-07 | 3.01377E-05 | 0.200908588 | 17.61107706 |
| 21 | 4786718.805 | 4.72134E-07 | 3.01428E-05 | 0.200945765 | 17.60629691 |
| 22 | 4644816.883 | 2.91281E-07 | 3.0223E-05 | 0.202151528 | 15.22202555 |
| 23 | 4631389.663 | 1.59444E-07 | 3.19886E-05 | 0.21450879 | 11.04862378 |
| 24 | 4489600.117 | 4.08188E-07 | 3.0161E-05 | 0.201304089 | 16.75923021 |
| 25 | 4382953.547 | 1.29189E-07 | 3.14197E-05 | 0.210795469 | 9.929168929 |
| 26 | 4356142.616 | 2.50723E-07 | 3.04739E-05 | 0.203987862 | 13.39663506 |
| 27 | 4355250.034 | 2.49254E-07 | 3.04727E-05 | 0.203985337 | 13.3809039 |
| 28 | 4335487.967 | 6.41115E-07 | 3.01108E-05 | 0.20010822 | 19.81656894 |
| 29 | 4226762.574 | 1.49214E-07 | 3.0556E-05 | 0.204913454 | 11.90770844 |
| 30 | 4115691.505 | 1.38846E-07 | 3.14839E-05 | 0.211191222 | 7.459834321 |
| 31 | 4072679.228 | 3.04457E-07 | 3.02154E-05 | 0.202051564 | 15.3793287 |
| 32 | 3932372.704 | 1.97705E-07 | 3.12901E-05 | 0.20967163 | 11.38121286 |
| 33 | 3864611.051 | 1.44803E-07 | 3.29215E-05 | 0.220835671 | 6.410976901 |
| 34 | 3750612.443 | 7.61273E-07 | 3.00531E-05 | 0.199277731 | 21.383684 |
| 35 | 3722327.172 | 1.45592E-07 | 3.28916E-05 | 0.220632017 | 6.897671194 |
| 36 | 3684853.908 | 2.15771E-07 | 3.07912E-05 | 0.206250338 | 11.69551341 |
| 37 | 3656307.745 | 1.3168E-07 | 3.3055E-05 | 0.221781772 | 6.156119184 |
| 38 | 3581084.462 | 1.45417E-07 | 3.29013E-05 | 0.220697725 | 6.420691259 |
| 39 | 3574102.037 | 1.46871E-07 | 3.29737E-05 | 0.221179676 | 6.37781606 |
| 40 | 3557449.533 | 1.58372E-07 | 4.27587E-05 | 0.286932892 | 0.578033365 |
| 41 | 3528699.024 | 1.55227E-07 | 4.31002E-05 | 0.289240448 | 0.525940059 |
| 42 | 3438917.589 | 1.41936E-07 | 3.09542E-05 | 0.207618275 | 10.12600856 |
| 43 | 3395880.46 | 3.46861E-07 | 3.02316E-05 | 0.202004487 | 15.84964087 |
| 44 | 3297992.633 | 3.60964E-07 | 3.02096E-05 | 0.201804876 | 16.00653593 |
| 45 | 3180446.649 | 1.49063E-07 | 3.09164E-05 | 0.207337895 | 10.14735474 |
| 46 | 3106722.218 | 1.52235E-07 | 3.77857E-05 | 0.253516331 | 2.904691725 |
| 47 | 3105496.908 | 1.52329E-07 | 3.77937E-05 | 0.253570016 | 2.898102931 |
| 48 | 3091881.862 | 1.53203E-07 | 3.90273E-05 | 0.261861129 | 1.603918416 |
| 49 | 3091184.09 | 1.531E-07 | 3.90172E-05 | 0.261794146 | 1.602770585 |
| 50 | 3021738.327 | 1.53389E-07 | 3.98348E-05 | 0.267290729 | 1.558477405 |
| 51 | 2885339.472 | 1.24338E-07 | 3.10629E-05 | 0.208414042 | 9.890222128 |
| 52 | 2874645.853 | 1.53621E-07 | 3.95888E-05 | 0.265635628 | 1.599423732 |
| 53 | 2854310.334 | 1.53972E-07 | 3.97016E-05 | 0.266392578 | 1.597202194 |
| 54 | 2847066.093 | 1.27112E-07 | 3.21436E-05 | 0.215670227 | 9.060510461 |
| 55 | 2630861.514 | 1.55004E-07 | 3.89627E-05 | 0.261420543 | 1.648844281 |
| 56 | 2628666.98 | 1.55845E-07 | 4.11665E-05 | 0.276236039 | 0.939751461 |
| 57 | 2564768.745 | 1.54999E-07 | 3.98216E-05 | 0.267195862 | 1.595499441 |
| 58 | 2526834.695 | 1.47958E-07 | 3.59066E-05 | 0.240896445 | 4.608506307 |
| 59 | 2504530.035 | 1.98914E-07 | 3.08089E-05 | 0.20643128 | 11.16119865 |
| 60 | 2460826.784 | 1.44548E-07 | 3.85403E-05 | 0.258618562 | 0.55303372 |
| 61 | 2439790.788 | 1.51158E-07 | 3.59447E-05 | 0.241141129 | 4.408199018 |
| 62 | 2414450.868 | 1.47132E-07 | 3.62487E-05 | 0.243200187 | 3.780157227 |
| 63 | 2411669.526 | 1.53714E-07 | 3.61916E-05 | 0.242792138 | 3.891025685 |
| 64 | 2387196.685 | 1.49959E-07 | 3.58956E-05 | 0.240815222 | 3.88391907 |
| 65 | 2351119.409 | 1.50929E-07 | 3.84824E-05 | 0.258205951 | 2.303531269 |
| 66 | 2268835.002 | 1.40145E-07 | 3.39013E-05 | 0.227441202 | 6.011963607 |
| 67 | 2213268.819 | 1.30379E-07 | 3.25004E-05 | 0.218057497 | 6.633146689 |
| 68 | 1946556.668 | 1.48144E-07 | 3.42886E-05 | 0.230016548 | 5.347714099 |
| 69 | 1920519.321 | 1.47141E-07 | 3.48246E-05 | 0.233623946 | 5.511532054 |
| 70 | 1862154.541 | 1.30472E-07 | 3.13628E-05 | 0.210408006 | 7.562302191 |
| 71 | 1837622.413 | 1.49263E-07 | 3.76746E-05 | 0.252780066 | 2.877145575 |
| 72 | 1827261.668 | 1.51106E-07 | 3.72667E-05 | 0.250030386 | 3.010138495 |
| 73 | 1797392.538 | 1.51307E-07 | 3.74982E-05 | 0.251586181 | 2.963300627 |
| 74 | 1678950.743 | 1.52483E-07 | 3.7474E-05 | 0.251419743 | 2.916741545 |

Table A.3  Machine Tool Selection Dataset

| | Alternatives | CC | SS | TC | TX | TZ | MD | ML |
|---|---|---|---|---|---|---|---|---|
| | | Min | Max | Max | Min | Min | Max | Max |
| 1 | YANG ML-25A | 1,550,000 | 3,465 | 8 | 20 | 20 | 280 | 520 |
| 2 | YCM TC-15 | 1,400,000 | 5,950 | 12 | 15 | 20 | 250 | 469 |
| 3 | VTURN 16 | 1,100,000 | 5,940 | 12 | 12 | 15 | 230 | 600 |
| 4 | FEMCO WNCL-20 | 1,500,000 | 3,465 | 12 | 6 | 12 | 260 | 420 |
| 5 | FEMCO WNCL-30 | 2,600,000 | 3,960 | 12 | 12 | 16 | 300 | 625 |
| 6 | EX-106 | 1,320,000 | 4,950 | 12 | 24 | 30 | 240 | 340 |
| 7 | ECOCA SJ20 | 1,180,000 | 4,480 | 8 | 24 | 24 | 250 | 330 |
| 8 | ECOCA SJ25 | 1,550,000 | 3,950 | 12 | 15 | 20 | 280 | 460 |
| 9 | TOPPER TNL-85A | 1,200,000 | 3,465 | 8 | 20 | 24 | 264 | 400 |
| 10 | TOPPER TNL-100AL | 1,400,000 | 2,970 | 12 | 24 | 30 | 300 | 600 |
| 11 | TOPPER TNL-85T | 1,350,000 | 3,465 | 12 | 30 | 30 | 264 | 350 |
| 12 | TOPPER TNL-100T | 1,450,000 | 2,970 | 12 | 20 | 24 | 300 | 400 |
| 13 | ATECH MT-52S | 1,376,000 | 4,752 | 12 | 20 | 24 | 235 | 350 |
| 14 | ATECH MT-52L | 1,440,000 | 4,752 | 12 | 20 | 24 | 235 | 600 |
| 15 | ATECH MT-75S | 1,824,000 | 3,790 | 10 | 12 | 20 | 300 | 530 |
| 16 | ATECH MT-75L | 1,920,000 | 3,790 | 10 | 12 | 20 | 300 | 1030 |

Table A.4 Non-Traditional Machining (NTM) Process Selection Dataset

| | Alternatives | TSF | PR | MRR | Cost | E | TF | TC | S | M | F |
|---|---|---|---|---|---|---|---|---|---|---|---|
| | | Min | Min | Max | Min | Max | Min | Min | Max | Max | Max |
| 1 | Ultrasonic machining | 1 | 10 | 500 | 2 | 4 | 2 | 3 | 1 | 5 | 5 |
| 2 | Water jet machining | 2.5 | 0.22 | 0.8 | 1 | 4 | 2 | 2 | 3 | 5 | 4 |
| 3 | Electrochemical machining | 3 | 100 | 400 | 5 | 2 | 3 | 1 | 3 | 1 | 1 |
| 4 | Chemical machining | 3 | 0.4 | 15 | 3 | 3 | 2 | 1 | 3 | 3 | 1 |
| 5 | Electric discharge machining | 3.5 | 2.7 | 800 | 3 | 4 | 4 | 4 | 3 | 1 | 5 |
| 6 | Wire electrical discharge machining | 3.5 | 2.5 | 600 | 3 | 4 | 4 | 4 | 3 | 1 | 5 |
| 7 | Electron beam machining | 2.5 | 0.2 | 1.6 | 4 | 5 | 2 | 1 | 3 | 5 | 5 |
| 8 | Laser beam machining | 2 | 1.4 | 0.1 | 3 | 5 | 2 | 1 | 3 | 5 | 5 |

**Exercises**

1. Perform normalization of the objective matrix for the fruit supply chain dataset in the Appendix, by sum, vector, max-min and max methods, whose equations are in Table 15.1. Compare the range of each normalized objective by each of the four normalization methods. (Answer: ranges are summarized in the following table.)

| Objective | Sum Method | Vector Method | Max-min Method | Max Method |
| --- | --- | --- | --- | --- |
| Cost | 0.0565 to 0.0724 | 0.225 to 0.289 | 0.0 to 1.0 | 0.780 to 1.0 |
| Responsiveness | 0.0588 to 0.0649 | 0.235 to 0.259 | 0.0 to 1.0 | 0.906 to 1.0 |

2. An objective matrix with 5 objectives (A, B, C, D and E) and 4 alternatives, is shown in the table below. The following parts can be answered using manual calculations or EMCDM445 program.

| Objective Names → | A (Min) | B (Max) | C (Max) | D (Min) | E (Min) |
| --- | --- | --- | --- | --- | --- |
| Alternatives ↓ | | | | | |
| 1 | 20 | 8 | 13 | 4 | 9 |
| 2 | 5 | 2 | 8 | 17 | 10 |
| 3 | 4 | 17 | 6 | 2 | 1 |
| 4 | 6 | 6 | 9 | 15 | 3 |

(a) Apply entropy weighting method to calculate the weights of the objectives. (Answer: weights for A, B, C, D and E are respectively 0.233374, 0.216633, 0.03823, 0.262035 and 0.249728)
(b) Use the weights calculated in part (a) and apply TOPSIS method to compute the closeness value for each alternative. Which alternative does TOPSIS recommend? (Answer for closeness of alternatives 1, 2, 3 and 4 are respectively 0.398895,

0.358287, 0.958917, 0.508982; TOPSIS recommends the alternative 3 with the largest closeness value)

(c) Use the weights you calculated in part (a) and apply PROBID method to compute the performance score for each alternative. Which alternative does PROBID recommend? (Answer for performance score of alternatives 1, 2, 3 and 4 are respectively 0.515521, 0.326647, 1.140223, 0.584214; PROBID recommends the alternative 3 with the largest performance score)

(d) Do TOPSIS and PROBID recommend the same alternative? Try other weighting and MCDM methods, and compare your results obtained in parts (b) and (c).

3. Use dataset given in the previous exercise and apply the following modifications to one or two objective functions.

(a) Linear transformation using y = 4x + 5. Then, perform MCDM on the modified dataset using entropy and CRITIC weighting methods (i.e., in two separate trials) and 5 MCDM methods. Compare the recommended alternatives with those based on the original dataset.

(b) Objective reformulation using y = 1/x, where x is the value of objective B. Then, perform MCDM on the modified dataset using entropy and CRITIC weighting methods (i.e., in two separate trials) and 5 MCDM methods. Compare the recommended alternatives with those based on the original dataset.

4. Using the EMCDM445 program, perform MCDM on fruit supply chain dataset in the Appendix, by all 5 MCDM methods with CRITIC weights and compare the recommended alternatives with those in Table 15.5. Repeat this with weights by BW method.

5. Using the EMCDM445 program, perform MCDM on VOC recovery dataset in the Appendix, by all 5 MCDM methods with the entropy weights and compare the recommended alternatives with those in Table 15.6. Repeat this with weights by AHP method.

6. Using the EMCDM445 program, perform MCDM on machine tool selection dataset in the Appendix, by all 5 MCDM methods with CRITIC weights and compare the recommended alternatives with those in Table 15.7. Repeat this with weights by AHP method.

7. Using the EMCDM445 program, perform MCDM on NTM process selection dataset in the Appendix, by all 5 MCDM methods with the entropy weights and compare the recommended alternatives with those in Table 15.8. Repeat this with weights by BW method.